\documentclass{article}

\usepackage[margin=1in]{geometry}
\usepackage{amsmath,amssymb}
\usepackage{graphicx}
\usepackage{hyperref}
\usepackage{natbib}
\usepackage{booktabs}
\usepackage{url}
\usepackage{authblk}
\usepackage{bm}

\newcommand{\amazonfn}{\footnote{Work does not relate to position at Amazon.}}

\title{JEPA as a Neural Tokenizer: Learning Robust Speech Representations with Density Adaptive Attention}

\author[1]{Georgios Ioannides\amazonfn}
\author[2]{Christos Constantinou\amazonfn}
\author[3]{Aman Chadha\amazonfn}
\author[4]{Aaron Elkins}
\author[5]{Linsey Pang}
\author[6]{Ravid Shwartz-Ziv}
\author[6]{Yann LeCun}

\affil[1]{Carnegie Mellon University, Amazon GenAI, James Silberrad Brown Center for Artificial Intelligence}
\affil[2]{University of Bristol, Amazon GenAI, James Silberrad Brown Center for Artificial Intelligence}
\affil[3]{Stanford University, Amazon GenAI, James Silberrad Brown Center for Artificial Intelligence}
\affil[4]{James Silberrad Brown Center for Artificial Intelligence}
\affil[5]{Northeastern University}
\affil[6]{New York University}

\date{October 25, 2025}

\begin{document}

\maketitle

\begin{abstract}
We introduce a two-stage self-supervised framework that combines the Joint-Embedding Predictive Architecture (JEPA) with a Density Adaptive Attention Mechanism (DAAM) for learning robust speech representations. Stage~1 uses JEPA with DAAM to learn semantic audio features via masked prediction in latent space, fully decoupled from waveform reconstruction. Stage~2 leverages these representations for efficient tokenization using Finite Scalar Quantization (FSQ) and a mixed-radix packing scheme, followed by high-fidelity waveform reconstruction with a HiFi-GAN decoder. By integrating Gaussian mixture-based density-adaptive gating into the JEPA encoder, the model performs adaptive temporal feature selection and discovers hierarchical speech structure at a low frame rate of 2.5~Hz. The resulting tokens (47.5 tokens/sec) provide a reversible, highly compressed, and language-model-friendly representation that is competitive with, and often more efficient than, existing neural audio codecs.
\end{abstract}

\tableofcontents

\section{Hybrid Discrete-Continuous Speech Representations via JEPA with Density Adaptive Attention}

\subsection{Overview}

We introduce a two-stage self-supervised learning framework that combines the Joint-Embedding Predictive Architecture (JEPA) \citep{Assran2023IJEPA} with Density Adaptive Attention Mechanisms (DAAM) for learning robust speech representations. This approach decouples representation learning from reconstruction: Stage~1 employs JEPA with DAAM to learn semantic audio features through masked prediction, while Stage~2 leverages these representations for efficient tokenization via Finite Scalar Quantization (FSQ) \citep{Mentzer2023FSQ} and high-quality reconstruction through HiFi-GAN \citep{Kong2020HiFiGAN}.

\textbf{Key innovation.} By integrating Density Adaptive Attention-based gating (Gaussian Mixture gating) \citep{Ioannides2024DAAM} into the JEPA encoder, we achieve adaptive feature selection during self-supervised learning. Combined with a mixed-radix packing scheme, the learned representations capture hierarchical speech structure---due to progressive downsampling from layer to layer---at a low frame rate of 2.5~Hz, enabling efficient speech modeling without labeled data.

\subsection{Motivation: Why JEPA for Speech?}

Traditional speech codec training couples representation learning with reconstruction objectives, forcing the encoder to prioritize features that minimize waveform-level losses. This conflates two distinct goals:
\begin{enumerate}
    \item Learning semantically meaningful representations that capture linguistic and acoustic structure.
    \item Preserving perceptual quality for high-fidelity reconstruction.
\end{enumerate}

JEPA addresses this by separating concerns: the encoder learns to predict masked representations in latent space (Stage~1), then a separate decoder learns to map these representations to audio (Stage~2). This architectural separation enables:
\begin{itemize}
    \item \textbf{Better representations:} the encoder optimizes for semantic content rather than low-level waveform details.
    \item \textbf{Efficiency:} fine-tuning the encoder reduces Stage~2 training cost.
    \item \textbf{Flexibility:} the same encoder can support multiple downstream tasks (text-to-speech, voice conversion, automatic speech recognition, etc.).
    \item \textbf{Scalability:} Stage~1 can leverage large unlabeled datasets.
\end{itemize}

The integration of DAAM enhances this framework by introducing adaptive attention that learns which temporal regions and features are most informative for prediction, naturally discovering speech-relevant patterns.

\section{Stage 1: Self-Supervised JEPA Encoder with DAAM}

\subsection{JEPA Masking Strategy}

The JEPA framework employs block-based temporal masking to create a self-supervised learning objective. For a batch of audio sequences with temporal length $T$, binary masks $\mathbf{m} \in \{0,1\}^{B \times T}$ are generated, where $1$ indicates visible (context) regions and $0$ indicates masked (target) regions.

\paragraph{Block Masking Algorithm.}

Given mask ratio $\rho \in [0,1]$, minimum span length $s_{\text{min}}$, and maximum span length $s_{\text{max}}$, we construct masks as follows:
\begin{enumerate}
    \item Initialize: $\mathbf{m} \leftarrow \mathbf{1}_{B \times T}$ (all positions visible).
    \item For each sample $b \in \{1, \ldots, B\}$:
    \begin{enumerate}
        \item Compute target: $n_{\text{mask}} = \lfloor \rho \cdot T \rfloor$.
        \item Initialize counter: $n_{\text{masked}} \leftarrow 0$.
    \end{enumerate}
    \item While $n_{\text{masked}} < n_{\text{mask}}$:
    \begin{enumerate}
        \item Sample span length: $\ell \sim \text{Uniform}(s_{\text{min}}, s_{\text{max}})$.
        \item Sample start position: $t_{\text{start}} \sim \text{Uniform}(0, T - \ell)$.
        \item Compute end position: $t_{\text{end}} \leftarrow \min(t_{\text{start}} + \ell, T)$.
        \item Set mask: $\mathbf{m}[b, t] \leftarrow 0$ for all $t \in [t_{\text{start}}, t_{\text{end}})$.
        \item Update counter: $n_{\text{masked}} \leftarrow n_{\text{masked}} + (t_{\text{end}} - t_{\text{start}})$.
    \end{enumerate}
    \item Return: mask tensor $\mathbf{m}$.
\end{enumerate}

This block masking strategy creates contiguous masked spans rather than random individual positions, forcing the model to learn longer-range temporal dependencies and semantic content.

\paragraph{Masking hyperparameters.}
\begin{itemize}
    \item Mask ratio: $\rho = 0.5$ (50\% of timesteps masked).
    \item Minimum span: $s_{\text{min}} = 2$ frames.
    \item Maximum span: $s_{\text{max}} = T/4$ frames (adaptive to sequence length).
\end{itemize}

At 2.5~Hz frame rate, this corresponds to variable spans adapted to the sequence length.

\begin{figure}[t]
    \centering    \includegraphics[width=0.6\linewidth]{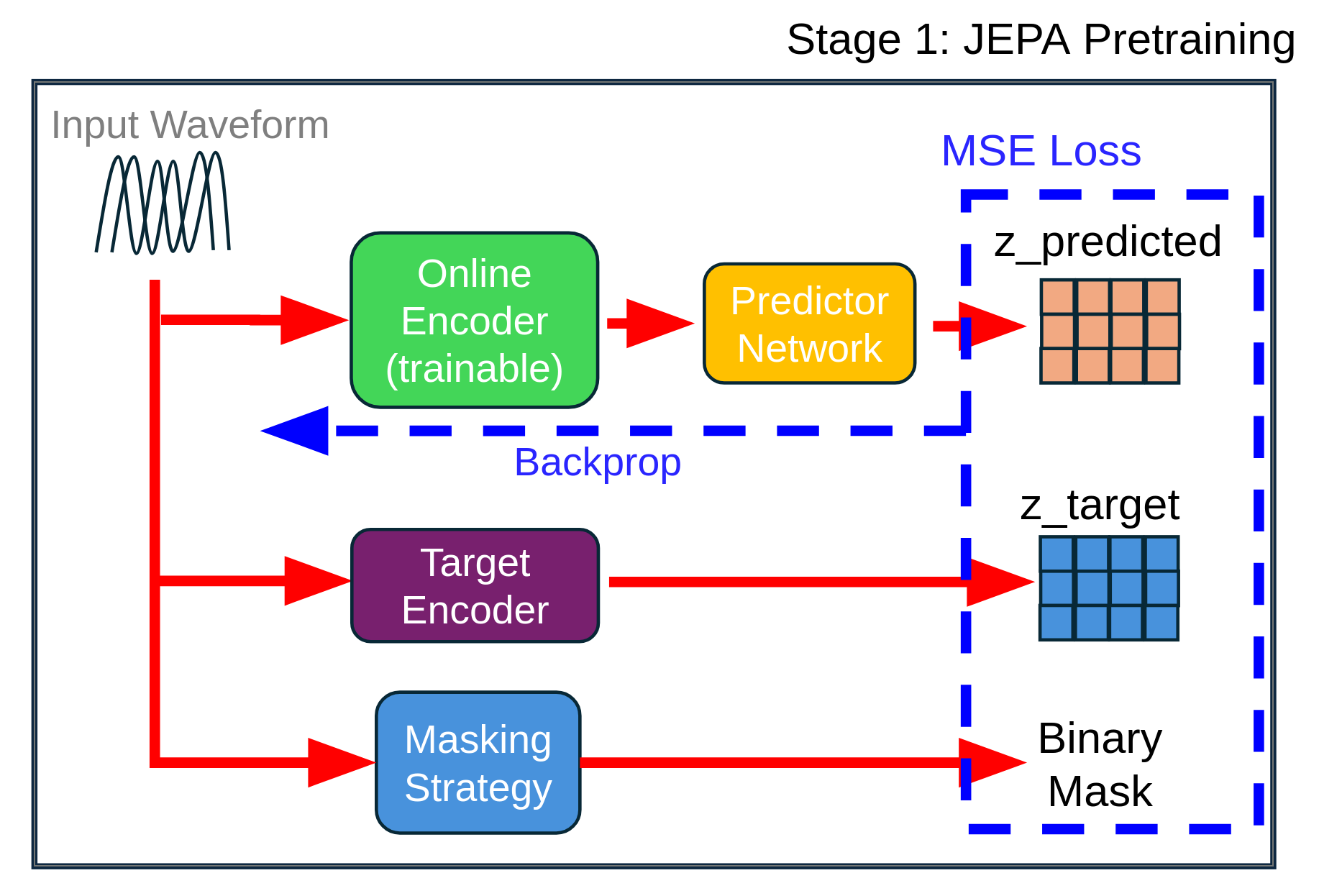}
    \caption{The input waveform is processed by three parallel pathways: (1) an online encoder (trainable, green) that processes the full audio and feeds into a predictor network (yellow) after feature-space masking with a learned mask token, (2) a target encoder (purple) updated via EMA that also processes the full audio to generate $\mathbf{z}_{\text{target}}$, and (3) a masking strategy module (blue) that generates binary masks. The MSE loss is computed only on masked regions between $\mathbf{z}_{\text{predicted}}$ and $\mathbf{z}_{\text{target}}$ (stop-gradient), with gradients backpropagating only through the online encoder and predictor. The target encoder provides stable representations without receiving gradients directly \citep{Grill2020BYOL}.}
    \label{fig:jepa-arch}
\end{figure}

\subsection{Density Adaptive Attention for Temporal Feature Modulation}

The core innovation integrating a stabilized version of the original DAAM into JEPA is the \emph{DensityAdaptiveAttention} module, which computes adaptive attention gates based on learned Gaussian mixture distributions. Unlike standard self-attention that computes pairwise dot-products between positions, DAAM learns to identify statistically salient temporal regions based on their distributional characteristics.

\subsubsection{Mathematical Formulation}

For input features $\mathbf{x} \in \mathbb{R}^{B \times C \times T}$ (batch size, channels, time), the DAAM module operates along the temporal axis.

\paragraph{Step 1: Temporal statistics.}

For each batch and channel, compute the mean and variance across time:
\begin{align}
\mu &= \frac{1}{T}\sum_{t=1}^T x_{:,:,t} \in \mathbb{R}^{B \times C \times 1}, \\
\sigma^2 &= \frac{1}{T}\sum_{t=1}^T (x_{:,:,t} - \mu)^2 \in \mathbb{R}^{B \times C \times 1}.
\end{align}

\paragraph{Step 2: Learnable Gaussian parameters.}

For $K$ Gaussian components, we maintain learnable parameters:
\begin{itemize}
    \item Mean offsets: $\boldsymbol{\delta} = [\delta_1, \ldots, \delta_K] \in \mathbb{R}^K$, initialized to $\delta_k = 0$.
    \item Log-scale parameters: $\boldsymbol{\nu} = [\nu_1, \ldots, \nu_K] \in \mathbb{R}^K$, initialized to $\nu_k = \log(0.5)$.
\end{itemize}

The positive scales are computed via softplus:
\begin{equation}
\tilde{\sigma}_k = \text{softplus}(\nu_k) + \epsilon = \log(1 + \exp(\nu_k)) + \epsilon,
\end{equation}
with $\epsilon = 10^{-3}$ for numerical stability.

\paragraph{Step 3: Standardized deviations.}

For each component $k$ and timestep $t$:
\begin{equation}
z_{k,t} = \frac{x_{:,:,t} - (\mu + \delta_k)}{\sigma \cdot \tilde{\sigma}_k + \epsilon}.
\end{equation}

\paragraph{Step 4: Log-density under each Gaussian.}

The log-probability density at each timestep is:
\begin{equation}
\log p_k(x_t) = -\frac{1}{2}z_{k,t}^2 - \log \tilde{\sigma}_k - \frac{1}{2}\log(2\pi).
\end{equation}

\paragraph{Step 5: Mixture aggregation via log-sum-exp.}

To form a mixture of Gaussians:
\begin{equation}
\log \mathbf{G}(x_t) =
\text{logsumexp}(\{\log p_1(x_t), \ldots, \log p_K(x_t)\}) - \log K.
\end{equation}

\paragraph{Step 6: Attention gate and feature modulation.}

The final attention gate is
\begin{equation}
\mathbf{G}(x_t) = \exp(\log \mathbf{G}(x_t)),
\end{equation}
and the output features are
\begin{equation}
\mathbf{y}_t = \mathbf{x}_t \odot \mathbf{G}(x_t),
\end{equation}
where $\odot$ denotes element-wise multiplication.

DAAM operates on a learned 1-channel attention projection over time: features are first projected to a single channel, the Gaussian mixture gate is computed on that 1D temporal signal, and the resulting gate scales the full feature tensor.

\paragraph{Implementation details.}
\begin{itemize}
    \item All computations in FP32 for numerical stability.
    \item Variance clamped: $\text{var} \geq 10^{-6}$.
    \item Softplus ensures positive scales: $\tilde{\sigma}_k > 0$.
    \item Number of Gaussians: $K = 4$ across all layers.
\end{itemize}

\subsection{JEPA Encoder Architecture}

The JEPA encoder consists of two parallel pathways that share weights but serve different roles.

\paragraph{Context encoder (online network).}
Processes the full audio input. Masking is applied later in feature space by replacing hidden timesteps with a learned mask token before the predictor. Parameters are updated via gradient descent.

\paragraph{Target encoder (EMA network).}
Processes the full audio input and provides stable targets for prediction. Parameters are updated via exponential moving average (EMA).

\subsubsection{Convolutional--Transformer Hybrid Design}

\paragraph{Downsampling path.}

The input raw waveform $[B, 1, T_{\text{wav}}]$ passes through Conv1D blocks with stride, progressing through channel dimensions
\[
64 \rightarrow 128 \rightarrow 256 \rightarrow 384 \rightarrow 512 \rightarrow 512.
\]
The total stride is $8 \times 8 \times 5 \times 5 \times 6 = 9600$ samples/hop at 24\,kHz, resulting in a latent representation $[B, 512, T_z]$, where $T_z$ corresponds to approximately 2.5~Hz frame rate.

\paragraph{Conformer blocks \citep{Gulati2020Conformer}.}
We use 8 Conformer layers with 16 attention heads. Each layer comprises self-attention, feedforward, convolution, and layer normalization. DAAM gating is applied in the encoder blocks (after the strided convolutions and residual stacks); there is no DAAM after the Conformer blocks in the current implementation.

\paragraph{Integration with DAAM.}

After each Conformer block, features pass through GAttnGateG modules that:
\begin{enumerate}
    \item Project features to a single channel via $1 \times 1$ convolution.
    \item Compute a DAAM gate from projected features.
    \item Apply learned scaling
    \begin{equation}
        \mathbf{y} = \mathbf{x} \cdot (1 + \alpha \cdot \text{gate}),
    \end{equation}
    where $\alpha$ (initialized to $0.05$) controls modulation strength.
\end{enumerate}

\begin{figure}[t]
    \centering
    \includegraphics[width=0.6\linewidth]{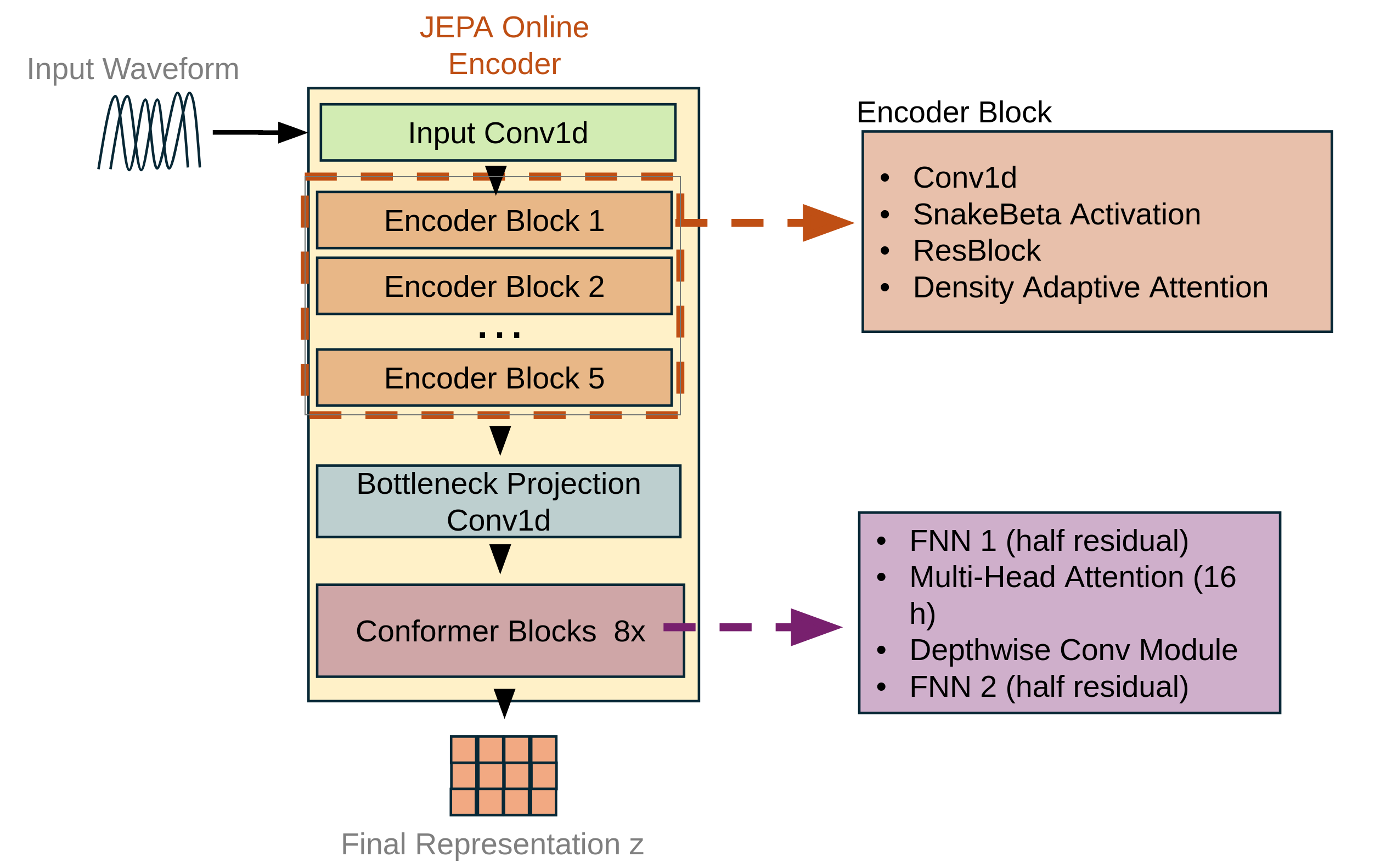}
    \caption{JEPA online encoder architecture. Input waveform passes through an initial Conv1D layer followed by 5 encoder blocks, each containing Conv1D with stride, SnakeBeta activation, residual blocks, and Gaussian Adaptive Attention gating. Features are projected through a bottleneck Conv1D layer and processed by 8 Conformer blocks (each with FNN, multi-head attention with 16 heads, depthwise convolution, and a second FNN) to produce the final representation $\mathbf{z}$. The target encoder shares this architecture but is updated via exponential moving average rather than backpropagation.}
    \label{fig:encoder}
\end{figure}

\subsection{JEPA Predictor Network}

The predictor takes context representations and predicts masked regions. It uses two Conformer blocks with 16 attention heads, processing masked context features and outputting predictions for all temporal positions. The predictor only receives context (visible) regions but must predict features at all positions; the mask is applied to the loss.

\begin{figure}[t]
    \centering
    \includegraphics[width=0.6\linewidth]{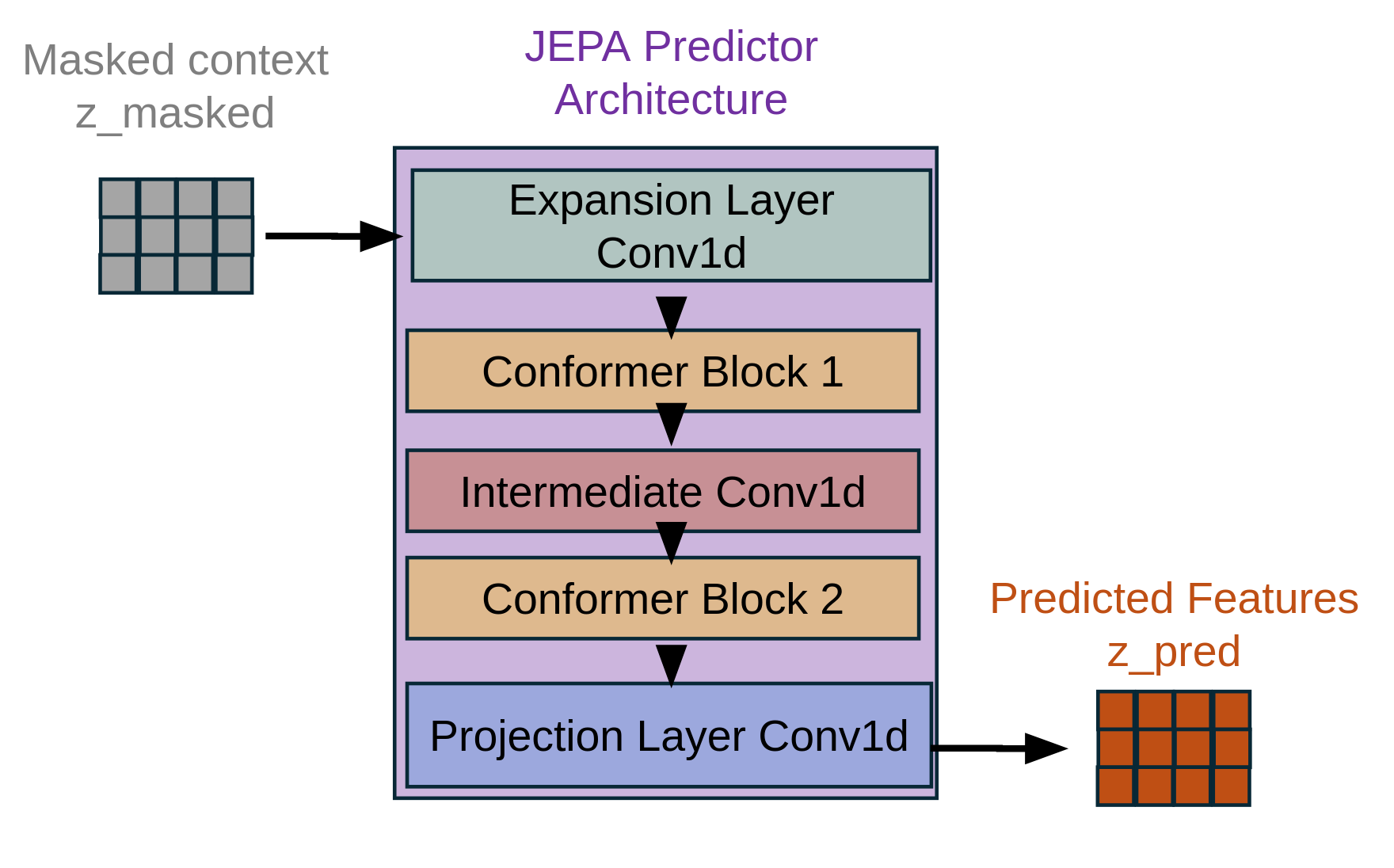}
    \caption{JEPA predictor network architecture. The predictor takes masked context features $\mathbf{z}_{\text{masked}}$ and processes them through: (1) an expansion Conv1D layer that doubles the channel dimension, (2) two Conformer blocks separated by an intermediate Conv1D for feature refinement, and (3) a projection Conv1D that reduces back to the original dimensionality, producing predicted features $\mathbf{z}_{\text{pred}}$ at all positions including masked regions.}
    \label{fig:predictor}
\end{figure}

\subsection{Stage 1 Training Objective}

The JEPA training objective is pure self-supervised prediction in latent space.

\subsubsection{Loss Function}

\begin{equation}
\mathcal{L}_{\text{JEPA}} =
\frac{1}{N_{\text{mask}} \cdot C}
\sum_{t \in \mathcal{M}}
\left\|
\mathbf{z}_{\text{pred}}^{(t)} -
\text{sg}(\mathbf{z}_{\text{target}}^{(t)})
\right\|^2,
\end{equation}
where:
\begin{itemize}
    \item $\mathcal{M} = \{t : m_t = 0\}$ is the set of masked positions,
    \item $N_{\text{mask}} = |\mathcal{M}|$,
    \item $C$ is the channel dimension,
    \item $\text{sg}(\cdot)$ denotes the stop-gradient operation.
\end{itemize}

The loss is computed only on masked regions by weighting squared differences and normalized by the number of masked tokens times channels.

\subsubsection{EMA Target Update}

After each training step, the target encoder parameters are updated via EMA:
\begin{equation}
\boldsymbol{\theta}_{\text{target}}
\leftarrow
\tau \boldsymbol{\theta}_{\text{target}}
+ (1-\tau)\boldsymbol{\theta}_{\text{online}},
\end{equation}
with momentum coefficient $\tau = 0.996$.

\paragraph{Stage 1 hyperparameters.}
\begin{itemize}
    \item Optimizer: AdamW with $\beta_1 = 0.8$, $\beta_2 = 0.99$.
    \item Learning rate: $1.5 \times 10^{-4}$.
    \item Weight decay: $10^{-3}$.
    \item Batch size: 32.
    \item Max audio length: 15\,s @ 24\,kHz.
    \item Training steps: 24\,000.
\end{itemize}

\paragraph{Collapse monitoring.}

We monitor (without backpropagation) the standard deviation of predictor outputs across batch and temporal dimensions. If the mean standard deviation falls below $0.01$, a warning is logged. This does not contribute to the loss.

\begin{figure}[t]
    \centering
    \includegraphics[width=0.6\linewidth]{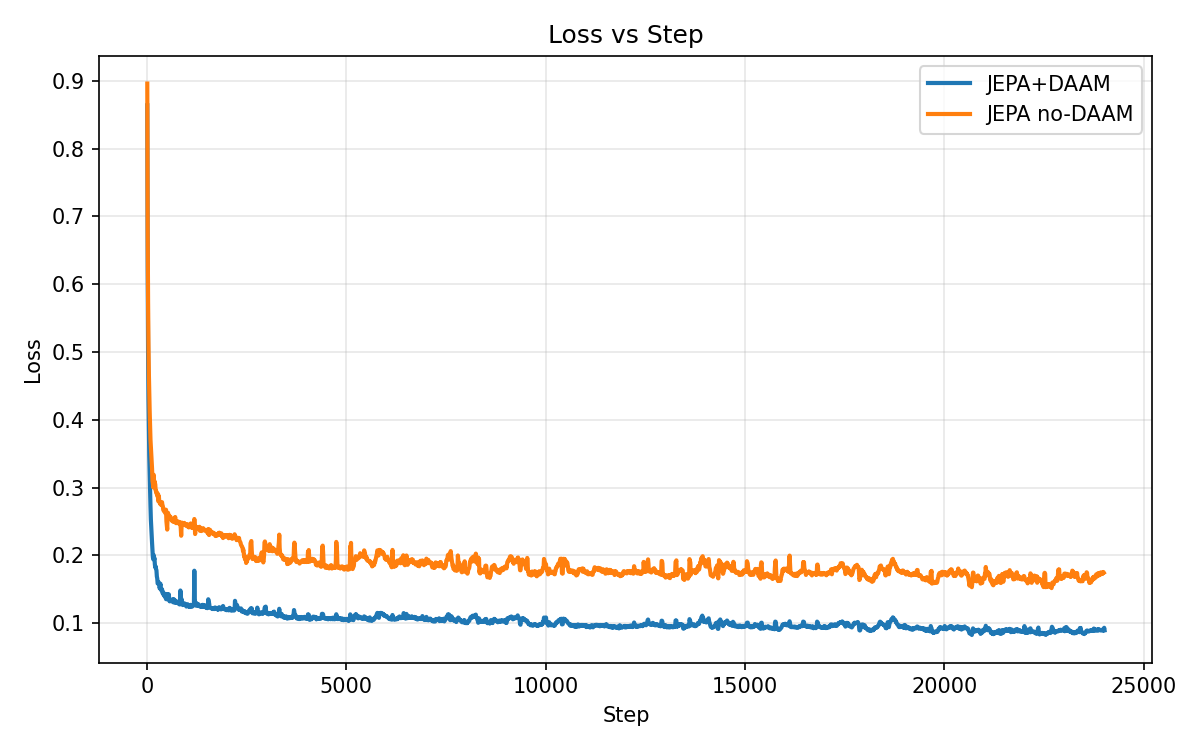}
    \caption{Stage~1 JEPA masked prediction loss (MSE) over training steps. JEPA+DAAM (blue) converges faster and to a lower final loss ($\sim 0.09$) compared to JEPA without DAAM (orange, $\sim 0.17$), demonstrating that Density Adaptive Attention enables more efficient representation learning. Both models use identical architectures except for DAAM gating.}
    \label{fig:loss}
\end{figure}

\section{Stage 2: Fine-Tuning Encoder + FSQ Quantization + HiFi-GAN Decoder}

After Stage~1 completes, the JEPA encoder weights are fine-tuned and used as a feature extractor for Stage~2. Stage~2 introduces quantization and waveform reconstruction.

\subsection{Finite Scalar Quantization (FSQ)}

FSQ provides efficient discrete tokenization without codebook learning \citep{Mentzer2023FSQ}. Unlike VQ-VAE, which maintains learnable codebooks, FSQ uses fixed scalar quantization per dimension.

Let $\mathbf{z}_e \in \mathbb{R}^{B \times C \times T}$ be encoder features.

\subsubsection{FSQ Formulation}

\paragraph{Projection.}
\begin{equation}
\mathbf{z}_e' = \tanh(\mathbf{z}_e).
\end{equation}

\paragraph{Quantization.}

For dimension $d$ with level $L_d$, define boundaries
\begin{equation}
B_d = \left\{ \frac{2i - L_d + 1}{L_d} : i \in \{0,1,\ldots,L_d-1\} \right\}.
\end{equation}
The quantization function is
\begin{equation}
q_d(x) = \arg\min_{b \in B_d} |x - b|.
\end{equation}
The quantized value is $\mathbf{z}_q[d] = q_d(\mathbf{z}_e'[d])$.

\paragraph{Configuration.}
\begin{itemize}
    \item Levels: $\mathbf{L} = [4,4,4,4]$.
    \item Code dimension: $C = 128$.
    \item Temperature: $\tau = 1.0$.
\end{itemize}

\paragraph{Straight-through estimator.}

During backpropagation,
\begin{equation}
\frac{\partial \mathcal{L}}{\partial \mathbf{z}_e}
=
\frac{\partial \mathcal{L}}{\partial \mathbf{z}_q}.
\end{equation}

\subsection{Mixed-Radix Token Packing}

To maximize compression efficiency, we implement a mixed-radix packing algorithm that converts FSQ indices into compact integer tokens \citep{Simon2024MixedRadixArxiv}.

Let $\mathbf{i} \in \mathbb{Z}^{B \times T \times D}$ denote FSQ indices, with dimension-specific radices $\mathbf{r} = [r_1,\ldots,r_G]$ for a group of $G$ dimensions.

\subsubsection{Mixed-Radix Encoding}

Any combination $[i_1,\ldots,i_G]$ is encoded as
\begin{equation}
\text{token}
= \sum_{k=1}^{G} i_k \prod_{j=k+1}^{G} r_j.
\end{equation}

\paragraph{Example.} For $G=7$ and $\mathbf{r} = [4,4,4,4,4,4,4]$ with $\mathbf{i} = [2,1,3,0,2,1,3]$:
\begin{align}
\text{token} &= 2 \cdot 4^6 + 1 \cdot 4^5 + 3 \cdot 4^4 + 0 \cdot 4^3 + 2 \cdot 4^2 + 1 \cdot 4^1 + 3 \cdot 4^0 \\
&= 10023,
\end{align}
with maximum value $4^7 - 1 = 16383$.

\subsubsection{Efficient Iterative Computation}

Using Horner's method \citep{MixedRadixKnuth1997}:
\begin{equation}
\text{token} = i_1 \cdot r_2 \cdots r_G + \cdots + i_{G-1}\cdot r_G + i_G,
\end{equation}
implemented right-to-left:
\begin{enumerate}
    \item Initialize $\text{token} = i_G$.
    \item For $k = G-1$ down to $1$:
    \[
    \text{token} = i_k + \text{token} \cdot r_k.
    \]
\end{enumerate}

\subsubsection{Padding and Grouping}

Our FSQ implementation yields $D=128$ quantized dimensions. We choose group size $G=7$:
\begin{itemize}
    \item Number of groups: $\lceil 128/7 \rceil = 19$.
    \item Padding: $19 \times 7 - 128 = 5$ dimensions with radix 1.
\end{itemize}

\paragraph{Token rate.}

Frame rate:
\[
f = \frac{\text{sample\_rate}}{\text{hop}} = \frac{24000}{9600} = 2.5~\text{Hz}.
\]
Groups per frame: $19$. Tokens/sec:
\[
\text{tps} = 2.5 \times 19 = 47.5.
\]

\subsubsection{Decoding}

The reverse operation extracts indices:
\begin{enumerate}
    \item Initialize $\text{rem} = \text{token}$.
    \item For $k = 1$ to $G$:
    \begin{itemize}
        \item $\text{prod} = \prod_{j=k+1}^{G} r_j$.
        \item $i_k = \left\lfloor \text{rem} / \text{prod} \right\rfloor$.
        \item $\text{rem} = \text{rem} \bmod \text{prod}$.
    \end{itemize}
\end{enumerate}

\subsubsection{Comparison to Alternatives}

\begin{table}[t]
\centering
\begin{tabular}{lccc}
\toprule
Approach & Tokens/sec & Reversible & Notes \\
\midrule
No packing (128 dims) & 320 & Yes & Each FSQ dim is a token \\
Mixed-radix (ours, $G=7$) & 47.5 & Yes & Pack 7 dims/token \\
VQ codebook & Variable & Yes & Requires learned codebook \\
\bottomrule
\end{tabular}
\caption{Comparison of tokenization approaches.}
\label{tab:tokenization}
\end{table}

Advantages:
\begin{itemize}
    \item Perfect reversibility via modular arithmetic.
    \item Near-optimal compression for given radices.
    \item No learned codebook (unlike VQ-VAE).
    \item Flexible grouping $G$ trading vocabulary size versus token rate.
    \item Integer-only operations, hardware-friendly.
\end{itemize}

With $G=7$ and radix 4, the per-token vocabulary is $4^7 = 16384$, comparable to subword vocabularies used in NLP.

\subsubsection{Integration with Language Models}

The compact tokens enable direct training of decoder-only Transformers for speech generation:
\begin{itemize}
    \item Input: discrete token sequence at 47.5 tokens/sec.
    \item Output: next-token prediction over a 16\,384-way vocabulary.
    \item Decoding: tokens $\rightarrow$ FSQ indices $\rightarrow$ dequantized features $\rightarrow$ waveform via HiFi-GAN.
\end{itemize}

\subsubsection{Frame Rate Comparison with Neural Codecs}

\begin{table}[t]
\centering
\begin{tabular}{lcl}
\toprule
Model & Frame Rate & Notes \\
\midrule
Ours (JEPA+FSQ) & 2.5~Hz & Mixed-radix packing (19 groups/frame) \\
U-Codec \citep{Yang2025UCodec} & 5~Hz & Ultra-low for LLM-TTS \\
Mimi \citep{LlamaMimi2025} & 12.5~Hz & Semantic distillation \\
DualCodec \citep{Li2025DualCodec} & 12.5--25~Hz & Dual-stream architecture \\
SoundStream (24\,kHz) \citep{Zeghidour2021SoundStream} & 75~Hz & 13.3\,ms frames \\
EnCodec (24\,kHz) \citep{Defossez2022EnCodec} & 75~Hz & 75 steps/sec @ 24\,kHz \\
DAC (44.1\,kHz) \citep{DACJAX2024TokenRate} & 86~Hz & Stride 512 @ 44.1\,kHz \\
\bottomrule
\end{tabular}
\caption{Frame rate comparison with state-of-the-art neural codecs.}
\label{tab:frame-rate}
\end{table}

\subsection{HiFi-GAN Decoder}

The decoder upsamples quantized representations back to waveform using HiFi-GAN with DAAM gating in residual blocks \citep{Kong2020HiFiGAN}.

\subsubsection{Decoder Architecture}

Quantized features $[B,512,T_z]$ are upsampled via ConvTranspose1D blocks through channel dimensions
\[
512 \rightarrow 384 \rightarrow 256 \rightarrow 128 \rightarrow 64,
\]
with strides $6,5,5,8,8$ (total stride $9600$), yielding output waveform $[B,1,T_{\text{wav}}]$.

Each block consists of:
\begin{itemize}
    \item Upsampling ConvTranspose1D.
    \item Multi-receptive-field (MRF) residual blocks with (optionally) DAAM gating.
\end{itemize}

\paragraph{ResBlock with DAAM.}

Each residual block contains:
\begin{enumerate}
    \item Leaky ReLU activation.
    \item Dilated convolution.
    \item Residual connection.
\end{enumerate}

\paragraph{Decoder hyperparameters.}
\begin{itemize}
    \item Upsample kernels: $[3,7,11,15,23,32]$.
    \item Residual blocks: 8 per stage.
\end{itemize}

\begin{figure}[t]
    \centering
    \includegraphics[width=0.6\linewidth]{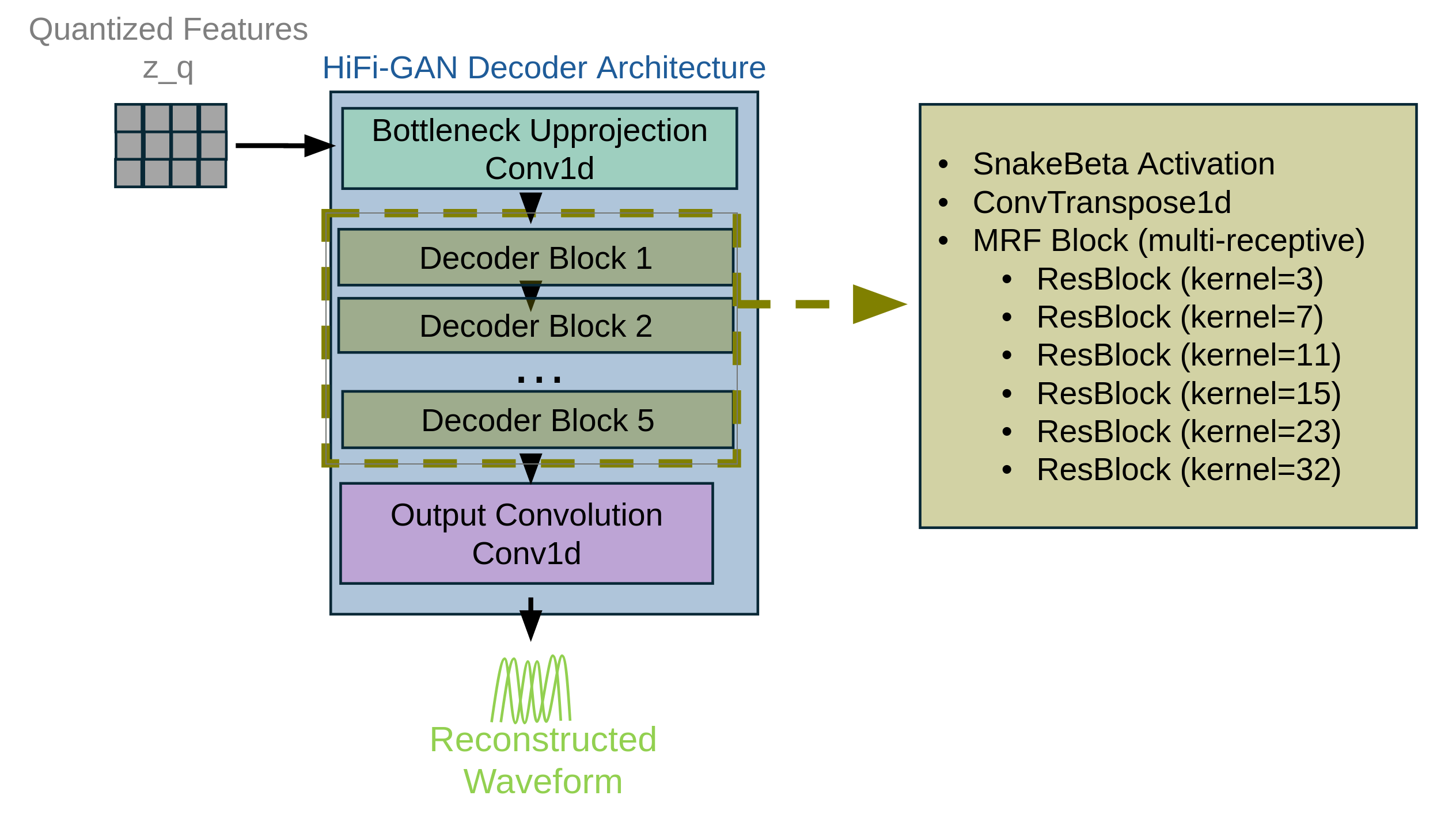}
    \caption{HiFi-GAN decoder architecture (Stage~2). Quantized features $\mathbf{z}_q$ are upsampled through a bottleneck Conv1D followed by 5 decoder blocks. Each block contains ConvTranspose1D upsampling and MRF residual blocks with different kernel sizes (3, 7, 11, 15, 23, 32) to capture multi-scale temporal patterns. SnakeBeta activations provide periodic inductive bias for high-fidelity audio generation \citep{Ziyin2020Snake}.}
    \label{fig:hifigan}
\end{figure}

\subsection{Stage 2 Training Objective}

Stage~2 optimizes the FSQ quantizer, HiFi-GAN decoder, and JEPA encoder.

\subsubsection{Total Loss}

\begin{equation}
\mathcal{L}_{\text{total}} =
\mathcal{L}_{\text{rec}} +
\lambda_{\text{stft}}\mathcal{L}_{\text{stft}} +
\lambda_{\text{gan}}\mathcal{L}_{\text{gan}}.
\end{equation}

\paragraph{1. Reconstruction loss (L1).}

\begin{equation}
\mathcal{L}_{\text{rec}} =
\frac{1}{T_{\text{wav}}}
\sum_{t=1}^{T_{\text{wav}}}
|\hat{x}_t - x_t|.
\end{equation}

\paragraph{2. Multi-resolution STFT loss \citep{Yamamoto2020ParallelWaveGAN}.}

\begin{equation}
\mathcal{L}_{\text{stft}} =
\sum_{m=1}^{M}
\left(
\mathcal{L}_{\text{sc}}^{(m)}
+
\mathcal{L}_{\text{mag}}^{(m)}
\right),
\end{equation}
with spectral convergence
\begin{equation}
\mathcal{L}_{\text{sc}}^{(m)} =
\frac{\left\|
|S_m(\hat{x})| - |S_m(x)|
\right\|_F}
{\left\| |S_m(x)| \right\|_F},
\end{equation}
and log-magnitude loss
\begin{equation}
\mathcal{L}_{\text{mag}}^{(m)} =
\frac{1}{N_m}
\left\|
\log |S_m(\hat{x})|
-
\log |S_m(x)|
\right\|_1.
\end{equation}

\paragraph{STFT configurations.}
\begin{itemize}
    \item FFT sizes: [2048, 1024, 512, 256, 128].
    \item Hop sizes: [512, 256, 128, 64, 32].
    \item Window: Hann.
\end{itemize}

\paragraph{3. GAN loss.}

We use multi-period and multi-scale discriminators \citep{Kumar2019MelGAN}.

Generator loss:
\begin{equation}
\mathcal{L}_{\text{gen}} =
\sum_{d \in \{\text{MPD}, \text{MSD}\}}
\mathbb{E}[(D_d(\hat{x}) - 1)^2].
\end{equation}

Feature matching:
\begin{equation}
\mathcal{L}_{\text{feat}} =
\sum_{d \in \{\text{MPD}, \text{MSD}\}}
\sum_{l=1}^{L_d}
\frac{1}{N_l}
\left\|
D_d^{(l)}(x) -
D_d^{(l)}(\hat{x})
\right\|_1.
\end{equation}

GAN total:
\begin{equation}
\mathcal{L}_{\text{gan}} =
\mathcal{L}_{\text{gen}} + \mathcal{L}_{\text{feat}}.
\end{equation}

Discriminator loss:
\begin{equation}
\mathcal{L}_{\text{disc}} =
\sum_{d \in \{\text{MPD}, \text{MSD}\}}
\left(
\mathbb{E}[(D_d(x) - 1)^2]
+
\mathbb{E}[D_d(\hat{x})^2]
\right).
\end{equation}

\paragraph{Loss weights and training schedule.}
\begin{itemize}
    \item $\lambda_{\text{stft}} = 2.0$.
    \item $\lambda_{\text{gan}} = 0.1$.
    \item Discriminator warmup: 5000 steps (disc frozen).
    \item After warmup: discriminator updated every step.
\end{itemize}

\paragraph{Stage~2 hyperparameters.}
\begin{itemize}
    \item Optimizer: AdamW, $\beta_1 = 0.8$, $\beta_2 = 0.99$.
    \item Learning rate: $1.5\times 10^{-4}$ (decoder), $0.75\times 10^{-4}$ (discriminators).
    \item Weight decay: $10^{-3}$.
    \item Batch size: 8.
    \item Training steps: 29\,000.
\end{itemize}

\section{Experimental Setup}

\subsection{Dataset}

\begin{itemize}
    \item LibriLight (large-scale unlabeled English speech corpus) \citep{Kahn2020LibriLight}.
    \item Training split: $\sim 9000$ hours (combined across the two stages).
    \item Validation: held-out speakers.
    \item Sample rate: 24\,kHz.
    \item Max audio length: 15\,s.
\end{itemize}

\subsection{Data Preprocessing}

\begin{enumerate}
    \item Resample to 24\,kHz if needed.
    \item Convert to mono by averaging channels.
    \item No further preprocessing (normalization handled in-model).
\end{enumerate}

\subsection{Distributed Training}

\begin{itemize}
    \item Hardware: 2x NVIDIA A100 (80\,GB).
    \item Mixed precision: FP16 for forward/backward, FP32 for critical ops.
    \item Gradient accumulation: 1 step.
    \item Global batch size: 64 (Stage~1), 16 (Stage~2).
\end{itemize}

\subsection{Inference Pipeline}

At inference time:
\begin{enumerate}
    \item Raw waveform $\rightarrow$ JEPA encoder $\rightarrow$ latent features.
    \item Latent features $\rightarrow$ FSQ quantization $\rightarrow$ discrete tokens.
    \item Tokens $\rightarrow$ dequantization $\rightarrow$ quantized features.
    \item Quantized features $\rightarrow$ HiFi-GAN decoder $\rightarrow$ reconstructed waveform.
\end{enumerate}

Token rate: 47.5 tokens/sec (with $G=7$ packing).

\section{Model Architecture and Efficiency}

\subsection{Parameter Counts}

\begin{table}[t]
\centering
\begin{tabular}{lcl}
\toprule
Component & Parameters & Notes \\
\midrule
\multicolumn{3}{l}{\textbf{Stage 1: JEPA encoder training}} \\
Online encoder & 121.7M & Trainable \\
Target encoder (EMA) & 118.5M & Momentum update \\
Predictor network & 3.2M & Trainable \\
Stage 1 total & \textbf{240.2M} & 121.7M trainable \\
\midrule
\multicolumn{3}{l}{\textbf{Stage 2: decoder training}} \\
JEPA encoder & 240.2M & Fine-tuned \\
FSQ quantizer & $\sim 0.01$M & Trainable \\
HiFi-GAN decoder & 69.2M & Trainable \\
Stage 2 total & \textbf{309.5M} & 69.3M trainable \\
\midrule
\multicolumn{3}{l}{\textbf{Final model (inference)}} \\
Encoder only & 121.7M & Online encoder only \\
FSQ + decoder & 69.3M & \\
Inference total & \textbf{191.0M} & Single-pass model \\
\bottomrule
\end{tabular}
\caption{Model architecture and parameter efficiency.}
\label{tab:params}
\end{table}

\subsection{Training Efficiency}

\begin{table}[t]
\centering
\begin{tabular}{lcc}
\toprule
Metric & Stage 1 (JEPA) & Stage 2 (decoder) \\
\midrule
Trainable parameters & 121.7M (50.7\%) & 69.3M (22.4\%) \\
Training steps & 24K & 29K \\
Batch size & 32 & 8 \\
Learning rate & $1.5\times 10^{-4}$ & $1.5\times 10^{-4}$ \\
\bottomrule
\end{tabular}
\caption{Training efficiency of the two stages.}
\label{tab:training-eff}
\end{table}

Key features:
\begin{itemize}
    \item Two-stage training: self-supervised pretraining + supervised fine-tuning.
    \item Inference efficiency: 191M parameters (no EMA network).
\end{itemize}

\section{Evaluation Metrics}

We report qualitative evaluations, as all variants were trained under limited computational budgets and this work presents preliminary findings.

\paragraph{Baselines.}
\begin{enumerate}
    \item JEPA baseline: JEPA encoder without DAAM gating.
    \item WavLM-Large \citep{Chen2021WavLM}: pre-trained self-supervised model.
    \item JEPA+DAAM: JEPA encoder with DAAM gating (ours).
\end{enumerate}

\section{Discussion}

\subsection{Why DAAM Improves JEPA Representations}

Integrating Density Adaptive Attention into JEPA provides several advantages.

\paragraph{Comparison to standard attention.}

Standard softmax-based self-attention computes pairwise correlations between positions, answering ``Which timesteps are similar to this one?'' DAAM instead computes statistical salience: ``Which timesteps have unusual or informative statistical properties?'' via Gaussian mixture modeling of temporal statistics.

Because it operates on temporal statistics rather than full pairwise similarity matrices, DAAM can capture salient temporal patterns without the quadratic complexity of full self-attention.

\section{Limitations and Future Work}

Current limitations and directions for future work include:
\begin{enumerate}
    \item \textbf{Fixed masking strategy.} Block masking with fixed span distributions may not adapt optimally to varying speech rates or linguistic structure. Future work includes adaptive masking sensitive to acoustic or linguistic boundaries.
    \item \textbf{Monolingual evaluation.} Experiments are currently limited to English (LibriLight). Generalization to tonal and morphologically rich languages remains open.
    \item \textbf{Limited data scale.} Pretraining has been conducted on relatively modest amounts of data compared to large-scale SSL systems; conclusions are restricted to emerging capabilities.
    \item \textbf{Cross-modal JEPA.} Extending to audio--visual or audio--text joint embedding prediction for multimodal representations is a promising direction.
\end{enumerate}

\section{Code Availability}

The complete implementation of the JEPA+DAAM framework, including training scripts, model architectures, and data processing pipelines, is available at:
\begin{center}
\url{https://github.com/gioannides/Density-Adaptive-JEPA}
\end{center}

The repository includes:
\begin{itemize}
    \item Stage~1 JEPA encoder training with DAAM.
    \item Stage~2 decoder training with the encoder.
    \item FSQ quantization and mixed-radix packing algorithms.
    \item HiFi-GAN decoder with optional DAAM gating.
    \item DeepSpeed integration for distributed training.
\end{itemize}

\section{Conclusion}

We introduced a two-stage self-supervised framework combining Joint-Embedding Predictive Architecture (JEPA) with Density Adaptive Attention Mechanisms (DAAM) for efficient speech representation learning. Stage~1 trains a JEPA encoder with DAAM-based gating to learn robust semantic representations via masked prediction using only MSE loss on masked regions. Stage~2 leverages these representations for reconstruction using L1 loss, multi-resolution STFT loss, and adversarial GAN losses, together with FSQ and HiFi-GAN.

Our main contributions are:
\begin{enumerate}
    \item A DAAM-enhanced JEPA encoder that uses Gaussian mixture-based attention for adaptive feature selection during self-supervised learning.
    \item An efficient tokenization scheme based on mixed-radix FSQ packing, achieving 47.5 tokens/sec, substantially lower than many existing neural audio codecs while remaining reversible.
    \item A two-stage training paradigm that cleanly separates representation learning from reconstruction, allowing pure self-supervised pretraining followed by reconstruction-focused fine-tuning.
\end{enumerate}

These results show that probabilistic attention mechanisms can improve representation learning by dynamically identifying acoustically salient regions during masked prediction, and that JEPA can serve as a powerful neural tokenizer for speech, suitable for integration with large language models and other sequence models.

\bibliographystyle{plainnat}
\bibliography{2025-10-25-jepa-daam}

\end{document}